\documentclass{elsart}

\usepackage{graphicx}
\usepackage{amsmath,amssymb}

\usepackage{natbib}

\begin{document}

\begin{frontmatter}
\title{Broad band Spectral Properties \\ of Accreting X--ray Binary Pulsars}
\author{Mauro Orlandini}
\address{INAF/IASF Bologna, via Gobetti 101, 40129 Bologna, Italy}

\begin{abstract}
Broad-band spectra of accreting X--ray binary pulsars can be fit by a
phenomenological model composed by a power law with a high energy rollover above
10 keV, plus a blackbody component with a temperature of few hundred eV. While,
at least qualitatively, the hard tail can be explained in terms of (inverse)
Compton scattering, the origin of the soft component cannot find a unique
explanation. Recently a qualitative picture able to explain the overall broad
band-spectrum of luminous X--ray pulsars was carried out by taking into account
the effect of bulk Comptonization in the accretion column. After a review on
these recent theoretical developments, I will present a case study of how
different modelization of the continuum affect broad features, in particular the
cyclotron resonance features in Vela X--1.
\end{abstract}


\end{frontmatter}

\section{Introduction}

\begin{figure}
\includegraphics[width=\textwidth]{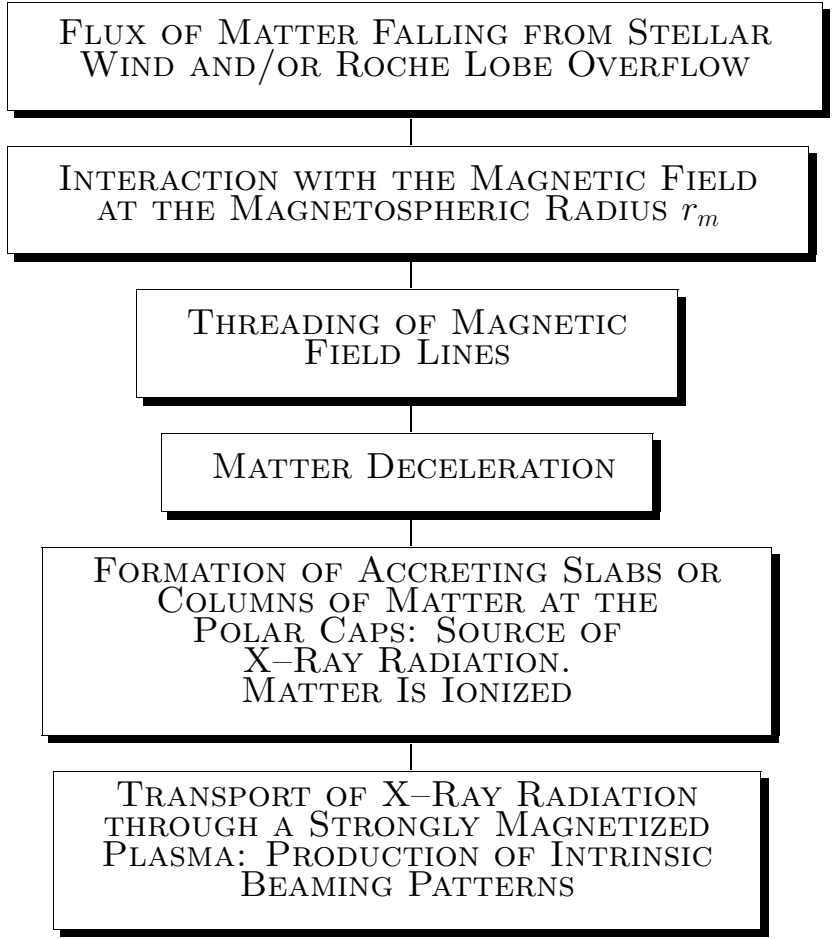}
\vspace{-4cm}
\caption[]{Schematic block diagram of the physical processes
occurring in a X--ray binary pulsar accreting from its companion. For a
detailed discussion on the physics involved in each block see \citet{2873}.}
\label{fig:block}
\end{figure}

Since the observation of pulsed emission from Centaurus X--3 \citep{1437} and the
discovery of its binary nature \citep{648}, more than 100 other X--ray binary
pulsars (XBPs) have been observed. The qualitative physical scenario able to
explain their pulsed X--ray emission was first elaborated by \citet{845} {\em
before} their discovery: X--rays are produced in the conversion into radiation of
the kinetic energy of matter coming from the stellar companion and accreting onto
the neutron star (NS). Because of the interaction with the NS strong magnetic
field, of the order of $10^{11}$--$10^{13}$~G, matter is driven onto the magnetic
polar caps, where it is decelerated and X--rays are produced. If the magnetic
field axis is not aligned with the spin axis, the NS acts as a ``lighthouse'',
giving rise to pulsed emission when the beam crosses our line of sight.  For a
detailed description of the XBP spectral properties it is therefore necessary to
describe the interactions of the X--rays produced at the NS surface with the
highly magnetized plasma forming the magnetosphere \citep[see, e.g.][]{1614}.  
This is a formidable task because we cannot use a linearized theory for the
radiative transfer equations but we have to deal with the fully
magnetohydrodynamical system. In Fig.~\ref{fig:block} we present a very schematic
block diagram of the physical scenario that describes the production and emission
of pulsed X-rays from XBPs \citep{2873}.

From the picture given above it should be clear that in order to be able to
describe the XBP spectra it is first necessary to understand how radiation is
formed and then to study how this radiation interacts with the strongly
magnetized plasma of the NS. As a consequence of this interaction, cyclotron
resonance features (CRFs) are observed in XRB spectra \citep[for a recent review
see][]{2673}. Because of their broad character, CRFs are quite sensible to the
continuum modelization. After a review on the mechanisms of spectral formation in
XBPs, I will present the case study of the Vela X--1 CRFs and their dependence on
the continuum adopted.

\section{Spectral formation mechanism in XBPs}

Once matter has penetrated the NS magnetosphere, it will follow the magnetic
field lines up to the magnetic polar cap, where it will be decelerated (see
Fig.~\ref{fig:block}). If the amount of matter falling onto the polar caps is
high enough that an X--ray luminosity greater than about $10^{37}$ erg/s is
reached, then a radiation-dominated (collisionless) shock will form at some
distance above the NS surface, creating an accretion column \citep{898}. For
luminosities lower than $10^{37}$ erg/s the radiation pressure is not sufficient
to stop the infalling matter, which therefore can impact directly onto the NS. In
this case the source of X--ray radiation is an accretion slab and emission occurs
in a direction parallel to the magnetic field (pencil beam), while in the case of
emission from an accretion column X--rays are emitted from the sides of the
column (fan beam) because the column is optically thick to X--ray radiation. It
is clear that the formation or not of an accretion column (that is, of a
collisionless shock) depends on how much X--ray radiation is produced which, in
turn, depends on the radiative transport through the infalling matter.  In other
words, the flow dynamics and the radiative transport are coupled, in a sort of
``feedback effect''.

From a theoretical point of view, the main physical mechanism of emission in XBPs
is (inverse) Compton scattering. According to the the value of the Comptonization
parameter $y$, we expect to observe a (modified) blackbody spectrum if $y\ll 1$,
while for $y\gg 1$ inverse Compton scattering can be important. If we define a
frequency $\omega_\text{co}$ such that $y(\omega_\text{co})=1$, then for
$\omega\gg\omega_\text{co}$ the inverse Compton scattering is saturated and the
emergent spectrum will show a Wien hump, due to low-energy photons up-scattered
up to $\hbar\omega\sim 3kT$ \citep{868}. In the case in which there is not
saturation a detailed analysis of the Kompaneets equation shows that the spectrum
will have the form of a power law modified by a high energy cutoff
\citep{868,614}.

Numerous attempts were made to numerically simulate XBP spectra \citep[see
e.g.][]{315,322}. While there was a qualitatively agreement with the
observations, an {\em ad hoc} source of soft photons was required in order to fit
the low energy part of the spectrum.  This low energy thermal emission was
discussed into detail by \citet{2986}. They find that, in the case of luminous
($\gtrsim10^{38}$ erg/s) pulsars, the most likely origin of this soft excess is
the reprocessing of hard X--rays (produced by the NS) by optically thick
material. This process can be excluded for less luminous sources
($\lesssim10^{36}$ erg/s), for which the most likely origin of the soft emission
is from the NS surface or circumstellar matter heated by the NS emission. For
intermediate luminosities XBPs are likely present both kind of emission.

Recently, \citet{3128,3181} renewed the interest on spectral formation in XBPs by
introducing, in the presence of an accretion column, the effect of bulk or
dynamical Comptonization. In the accretion column, the infalling electrons that
act as scattering medium for the soft photons possess a preferred motion: this
implies that photons gain energy through first-order Fermi acceleration. In the
case considered in the past of the thermal Comptonization process, photons gain
energy via second-order Fermi acceleration because of the incoherent, stochastic
motion of the plasma.  Therefore the energization process occurs at the expense
of the bulk kinetic energy of the infalling matter and is not supplied by the gas
internal energy. In this scenario the soft component comes out naturally from
soft photons produced by the ``thermal mound'' at the base of the accretion
column that are able to escape without experiencing many scatterings.

Taking into account {\em only} bulk Comptonization and neglecting cyclotron
emission and absorption, \citet{3181} were able to reproduce qualitatively the
general form of a XBP broad-band spectrum: they predict a high energy power law
and a turnover at low energy. This model is not able to explain the high energy
cutoff observed in many XBPs, but this is probably due to the not inclusion of
cyclotron processes: indeed the cutoff energy and the cyclotron resonance feature
(CRF) energy are strongly correlated \citep{407}.

\section{Comparison with observations}

\begin{figure}
\begin{center}
\includegraphics[height=1.1\textwidth,width=0.5\textheight,angle=270]{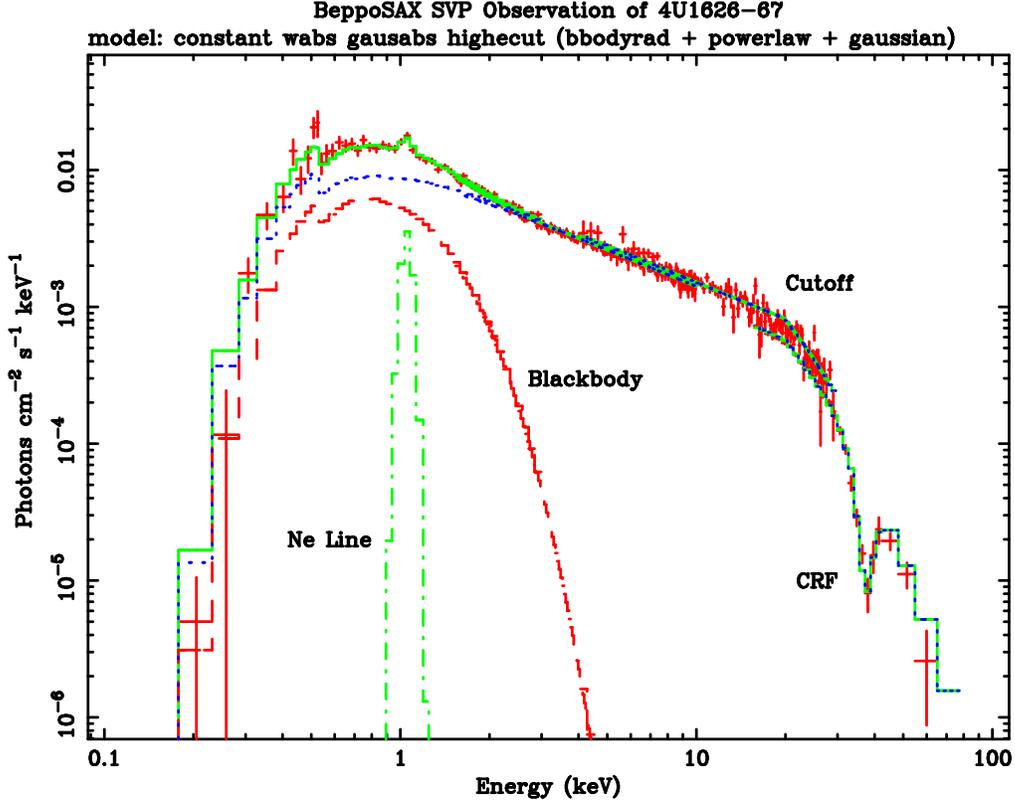}
\end{center}
\caption[]{0.1--100 keV pulse averaged spectrum of the XBP 4U~1626--67 as
observed by BeppoSAX \citep{1622}. All the typical XBP spectral components 
described in the text are present in this source.}
\label{fig:1626}
\end{figure}

From an observational point of view the phenomenological model able to describe
the broad-band spectra of accreting XBPs contains (i) a black-body component with
temperature of few hundreds eV; (ii) a power law of photon index $\sim$1 up to
$\sim$10~keV; and a (iii) a high energy ($\gtrsim 10$~keV) cutoff that makes the
spectrum rapidly drop above $\sim$40--50~keV \citep[see, e.g.][]{1502,2481,3258}.  
Superimposed there are emission line features due to fluorescence from ions at
different ionization levels, and broad absorption ``lines'' due to cyclotron
resonances. In Fig.~\ref{fig:1626} we show the BeppoSAX 0.1--100 keV, pulse
averaged spectrum of the XBP 4U~1626--67 \citep{1622}, in which all the
components mentioned above are clearly present.

The first description of the high energy spectral rollover was quite crude
\citep{303}, and \citet{1584} ``smoothed'' the cutoff by introducing the
so-called Fermi-Dirac cutoff. By analyzing a sample of XBP spectra observed by
{\em Ginga}, \citet{1754} introduced the so-called NPEX (Negative Positive
Exponential) model:

$$
\text{NPEX}(E) = (AE^{-\alpha} + BE^{+\beta}) \exp\left(-\frac{E}{kT}\right)
$$

the components of which have a physical meaning because, if $\beta=2$, it mimics
the saturated inverse Compton spectrum \citep[see discussion in][]{1502}.
Furthermore, because the (non relativistic) energy variation of a photon during
Compton scattering is $\Delta E/E = (4kT-E)/mc^2$ \citep{868} then when
$E=E_\text{cyc}$, with $E_\text{cyc}$ the CRF energy, the medium is optically
thick and therefore $E_\text{cyc}\sim 4kT$. In other words, the measurement of
the CRF energy gives an order of magnitude estimate of the temperature of the
electrons responsible of the resonance.

\subsection{The CRFs in Vela X--1}

\begin{figure}
\begin{center}
\includegraphics[height=1.1\textwidth,width=0.3\textheight,angle=270]{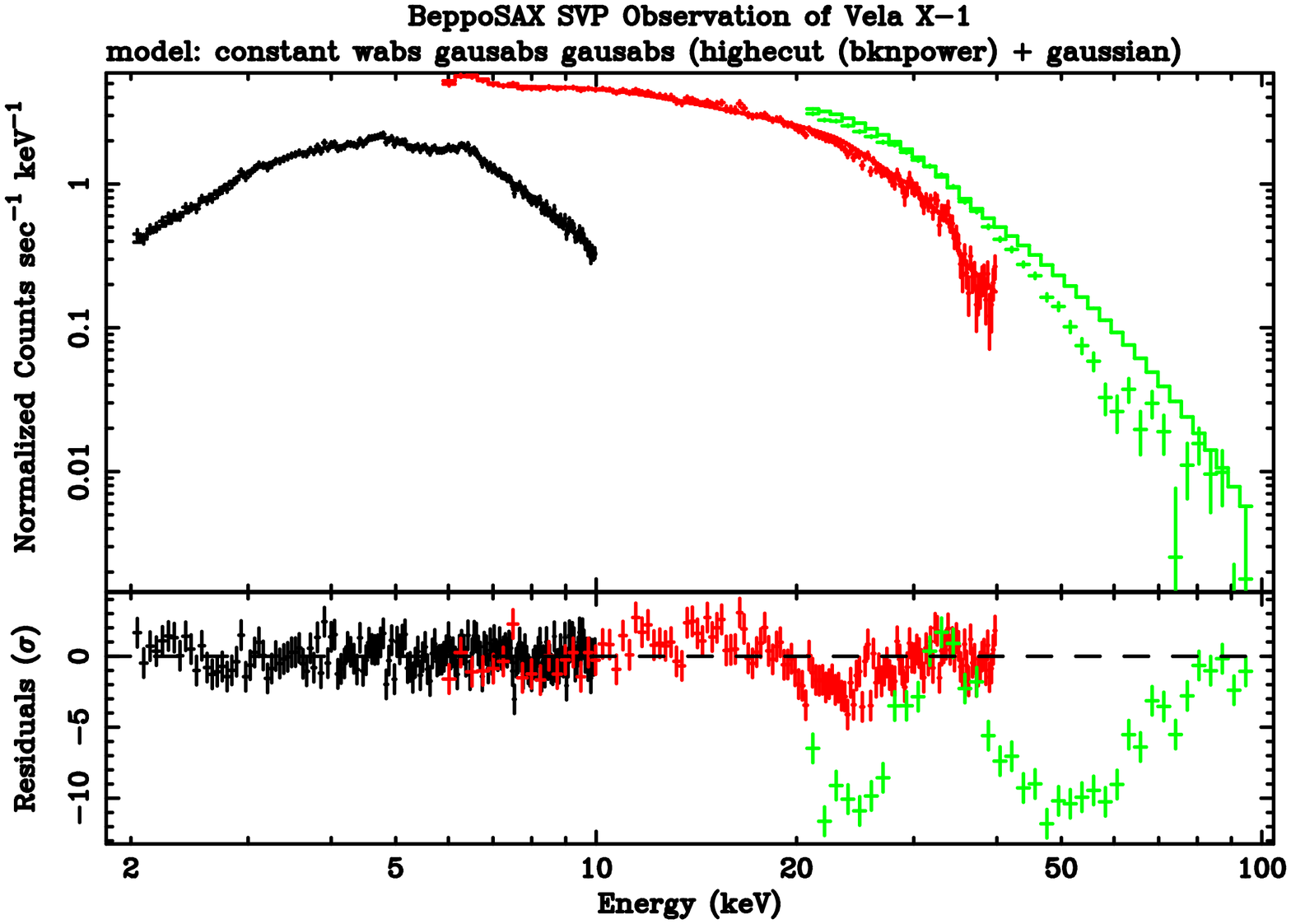}
\par\vspace{-0.5cm}
\includegraphics[height=1.1\textwidth,width=0.3\textheight,angle=270]{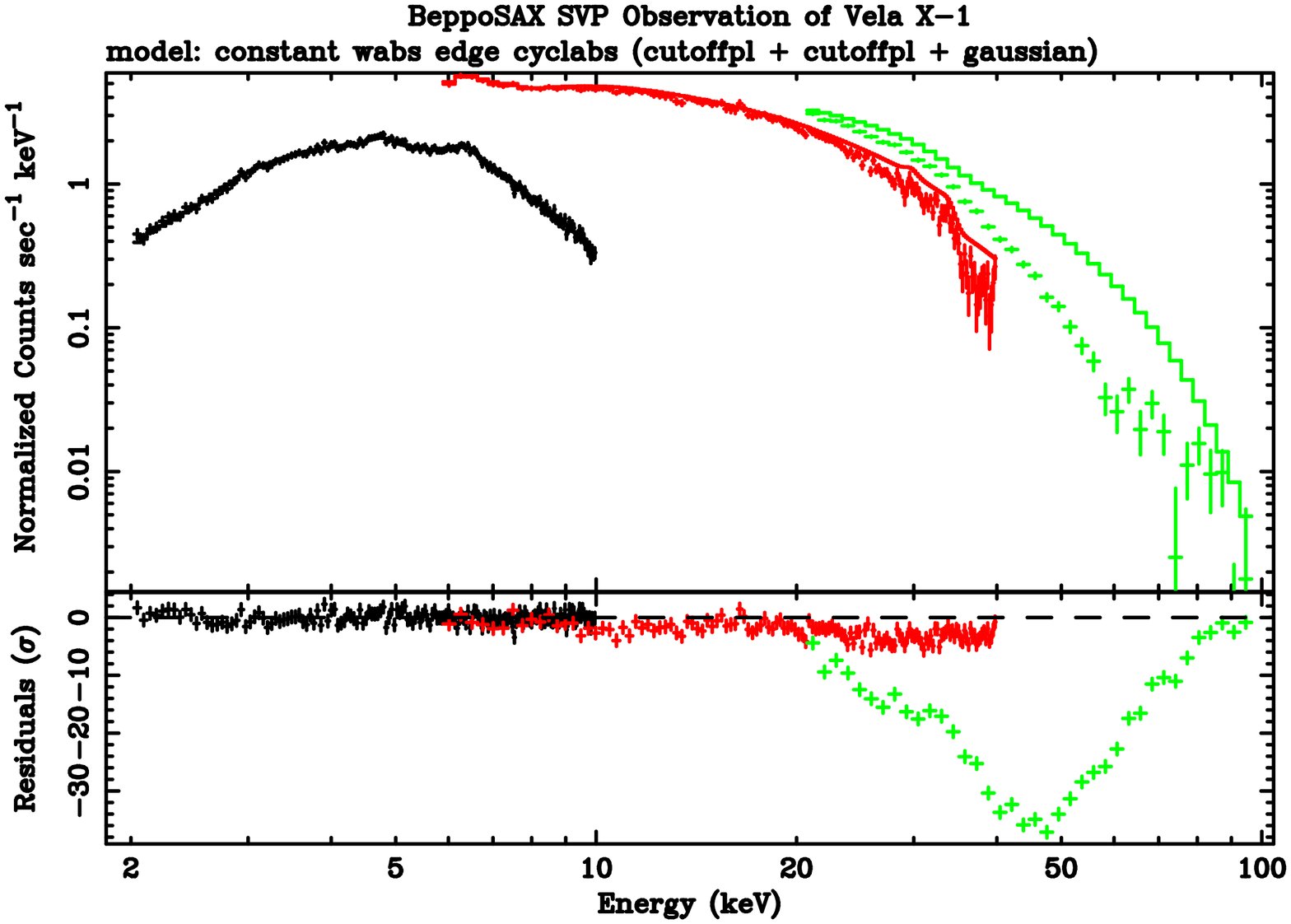}
\par\vspace{-0.5cm}
\includegraphics[height=1.1\textwidth,width=0.3\textheight,angle=270]{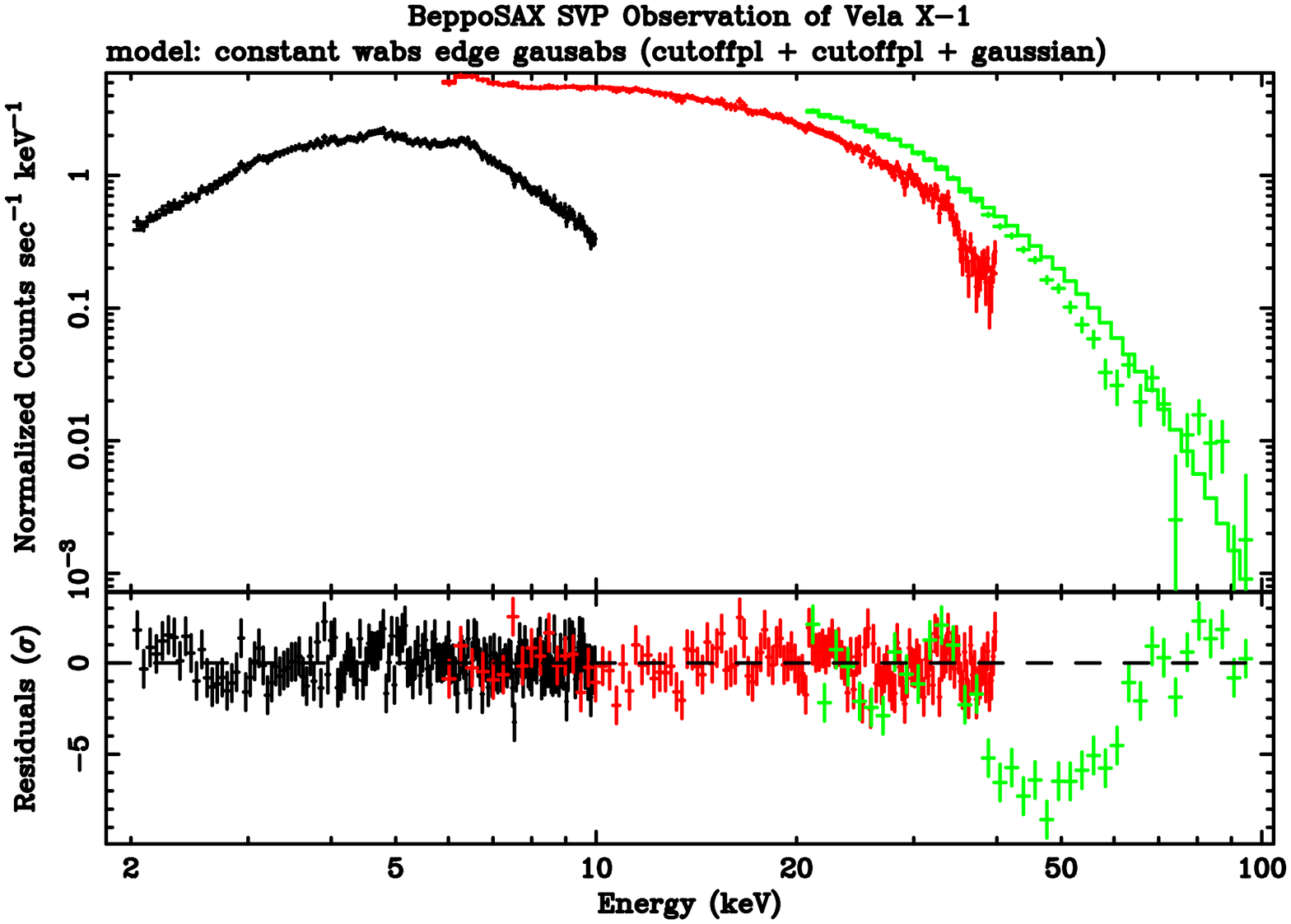}
\end{center}
\caption{BeppoSAX Vela X--1 broad-band spectrum fit with three different
spectral models \citep{1581}. The residual panels show the CRF profile obtained 
by putting the line normalization to zero. See text for the model descriptions.}
\label{fig:vela}
\end{figure}

The knowledge of the form of the continuum is of paramount importance for the
determination of the physical parameters derived from broad features, as CRFs. In
Fig.~\ref{fig:vela} we show the fits with different spectral models to the
BeppoSAX 0.1--100 keV, pulse-averaged spectrum of one of the best studied XBPs,
Vela X--1. This source presents a controversial CRF at $\sim25$ keV and a well
determined CRF at $\sim50$ keV \citep{1581,2519}. In order to check if the first
CRF is real or an artifact due to a incorrect modelization of the continuum, we
fit the spectrum with three different continuum models: a broken power law (top
panel in Fig.~\ref{fig:vela}), and two cutoff power laws (second and third
panel). In the first case we needed {\em two} CRFs, while in the other two cases
only one CRF at $\sim50$ keV was required. In the second and third panel we
modeled the CRF with the {\sc cycabs} model \citep{1754}, and a Gaussian in
absorption \citep{1172}, respectively. In order to extract the CRF profile from
the residual panel, we put to zero the CRF normalization. It is quite evident
from the residual panels in Fig.~\ref{fig:vela} that the CRFs strongly depend on
the choice of the continuum model (from an F--test they result statistically
equivalent): the $\sim25$ keV features is surely present with a broken power law
continuum (although its energy is quite close to the break energy)  while it
disappears if using two cutoff power laws. In this latter case there is a hint of
its presence by using a {\sc cycabs} modelization of the CRF, while it is not
present at all by using a Gaussian in absorption. A pulse phase spectroscopy
performed on a longer BeppoSAX observation confirmed this result \citep{2668}.  
We think that this feature could be due to a two-step steepening of the spectrum
modeled with an incorrect smooth rollover, as already observed in the XBP
OAO~1657--415 \citep{1961}.

\section{Conclusion}

After more that 30 years from their discovery, there is not yet a satisfactory
theoretical model for the spectral emission in XBPs able to explain the observed
spectra. Even if the recognition of the importance of bulk Comptonization is
opening new insight in the comprehension of this quite complicated issue, it is
the lack of a theoretical treatment of Compton scattering in the presence of
strong magnetic fields the likely culprit of our poor understanding of spectral
formation XBPs.


\end{document}